\documentclass[sigconf]{acmart}

\usepackage{booktabs}
\usepackage{cleveref}
\usepackage{dirtytalk}
\usepackage{natbib}
\usepackage{subfig}
\usepackage{enumitem}
\usepackage{url}
\usepackage{bbding} 
\usepackage{framed}

\begin{document}

\title{Agility is responsiveness to change: An essential definition}

\author{Lucas Gren}
\orcid{1234-5678-9012}
\affiliation{%
  \institution{Volvo Cars}
  \city{Gothenburg} 
  \state{Sweden} 
}
\email{lucas.gren@volvocars.com}

\author{Per Lenberg}
\affiliation{%
  \institution{Saab AB\\}
  \city{Gothenburg} 
  \state{Sweden} 
}
\email{per.lenberg@saabgroup.com}

\begin{abstract}
There is some ambiguity of what agile means in both research and practice. Authors have suggested a diversity of different definitions, through which it is difficult to interpret what agile really is. The concept, however, exists in its implementation through agile practices. In this vision paper, we argue that adopting an agile approach boils down to being more responsive to change. To support this claim, we relate agile principles, practices, the agile manifesto, and our own experiences to this core definition. We envision that agile transformations would be, and are, much easier using this definition and contextualizing its implications. 
\end{abstract}


\copyrightyear{2020}
\acmYear{2020}
\setcopyright{acmlicensed}
\acmConference[EASE 2020]{Evaluation and Assessment in Software Engineering}{April 15--17, 2020}{Trondheim, Norway}
\acmPrice{15.00}
\acmDOI{10.1145/3383219.3383265}
\acmISBN{978-1-4503-7731-7/20/04}


\maketitle

\section{Why agile software development is misunderstood}\label{sec:life}
We hardly need to provide an introduction to agility in the way we have done in all our research papers. Such an introduction would probably include the agile manifesto, its principles, and some descriptions of how the existing agile practices try to implement the principles. Agile has become a well-known concept. The concept, however, exists only in its implementation instances \citep{laanti2013definitions} and in success factors that vary \citep{chow2008survey}.

In this vision paper, we aim to define the core of the agile approach drawing on our experiences of six agile transformations with a hope that this would reshape the course of empirical agile development research and practice. The studied transformations were conducted in large (more than 5000 employees) organizations developing complex systems requiring multi-team collaboration (i.e.\ a scaled agile approach). The software systems developed by these companies were, however, not the sole delivery, but an essential component of a larger product. The experiences were collected over a period of more than ten years. In the pursuit of this definition, we wanted to distill how an agile paradigm might be different from all other popular paradigms in practice.

\citet{laanti2013definitions} listed all the definition of agility up until 2013 in research and they range from describing effectiveness, ability to steer, rule-base, people, communication, speed, flexibility, responsiveness, empowerment, change, feedback, value, delivery, innovations, adaptability, collaborative, iterative development, self-organization, light-weight process, cost-conscious, customer-driven, strategic, conceptual framework, and so on and so forth. Even the principles of the agile manifesto spans from customer satisfaction, continuous delivery, value, early delivery, adaptability, competitiveness, customer benefit, collaboration, motivated individuals, good environment, support, trust, efficiency, communication, progress measurement, sustainability, people, technical excellence, simplicity, optimization of work, self-organization, built in improvement of efficiency and behavior, and so on and so forth.

According to our observations, many practitioners are confused about what agile is. The existing agile definitions comprise all the studies done in organizational and social psychology, management, and engineering research in the last century, but without any references to these results. This is a pity because the agile ways-of-working work, and when companies see other companies succeeding with agile transformations, they also want to get on that (agile release) train.

Some define agile by comparing it to waterfall-like and plan-driven development methods. In a rigid structure where one phase of the large project needs to be finished before the next phase can start, being agile seems difficult, which it also is. However, many companies, and especially start-ups, had projects that were different even before agile became famous, and it is cumbersome to define a concept only by contrasting it to something else. 

We also argue that most, if not all, organizations tailor the agile construct, meaning that they interpret the definition so that it makes sense to their context. The problem is that, since the core of agile is not well defined, they often interpret agile as being a set of practices that conveniently fulfill their already existing political agendas. The type of activity chosen is often also focused on process- and technology-related aspects ignoring essential factors related to organizational values and social norms \citep{lenberg2}, probably due to the mentioned power struggles and politics. If a large company wants to be more agile, it needs as much focus on the cultural change as the technical infrastructure. So, what is at the core of agile and how is it different from how companies organized work before?

\section{Defining the core of agile}\label{sec:life}
Many people (including us) resort to Wikipedia to obtain a first idea of something new. Wikipedia cites \citet{Collier2011aaa} who defines agile software development as ``an approach to software development under which requirements and solutions evolve through the collaborative effort of self-organizing and cross-functional teams and their customer(s)/end user(s).'' It continues by citing the Agile Alliance web-page (https://www.agilealliance.org/agile101/) where agility is defined as ``the ability to create and respond to change in order to succeed in an uncertain and turbulent environment.'' We argue the latter is the best one so far, but that it could be shortened to ``responsiveness to change,'' and still be well understood. The reason why we do not want to go even farther and just use the definition in any dictionary that agile means to be flexible and to move quickly and easily, is that we then lose the development context of responding to some change external to the company itself. We have experienced that the term ``responsiveness to change'' is correctly interpreted in itself as external changes to the company in what adds customer value. Adapting to internal hierarchies and organizational politics are examples of changes that an agile company could adapt to, but such flexibility should not be the end goal, and are also not bundled into the term ``responsiveness to change'' in most agile transformations. We acknowledge that this definition is an incremental change, but the main point we are trying to make is to not lose this focus and get lost in scaled agile frameworks, which is what we see in practice.

We, moreover, argue that having a clear definition and understanding of the core concept is a necessity during the agile transformation (i.e.\ the organizational change). It facilitates the change process by giving it a distinct direction, enabling organizations to focus on what really matters. Most software organization are still hierarchical, and the management units a few steps away from the software engineering teams have little to no interest in, or knowledge of, process-oriented constructs like sprints, backlogs, or retrospectives. It is unlikely for a large organization to do a complete re-organization as part of their agile transformation; instead, the agile approach must work within a hierarchical structure where different units have different competencies, knowledge, and goals. To effectively communicate the value of agility with them, it must be simplified into something that has meaning for them specifically.

Without a common definition of what being agile means, interpretations may vary between various organizational units, which could cause miscommunication, confusion, or even conflict. If the construct is defined too broadly, the individual teams and organizational units will create their, more narrow, interpretations based on their needs, goals, and experiences. That, in turn, creates a misaligned organization. By forming a shared definition of what agile brings to the table, we minimize that risk.

The problem with dumping all aspects of modern organizational psychology into an approach is that too many employees in the transforming organizations will be confused. They will not, even after a long time, grasp what we consider to be the true meaning of agility and how it differs from what they used to do. Nor will they understand how that should affect their working methods. Furthermore, agile has become a hollow and general concept, and organizations affirm that they have a good (i.e.\ agile) enterprise or that they do good (i.e.\ agile) software development. We have seen that this causes organizations to adopt agile methods for the wrong reason, meaning that they do it because they have to, without in-depth insights into what it means. They want to be classified as an agile company, but they might not be fully committed to conducing the necessary changes.

Studies that look at practices have, for example, a hard time distinguishing between agile and lean \citep{isleanagile}. There are a lot of methods and practices on top of these two paradigms. The core of lean relates efficiency (i.e.\ doing things right), which is about removing waste while sustaining the same productivity \citep{Womack2003ltb}. Agile, however, we relate to effectiveness -- doing the right things. We, therefore, argue that agile is all about responsiveness to change. In the following section, we will relate most of the agile principles and methods to that core definition. We envision, and have observed, that by focusing on this simplified core statement, companies are more likely to have a successful agile transformation. Since organizational change requires much energy and can cause stress, companies need to keep their focus on what matters (i.e.\ what they can, and want, to achieve). When we state that agile is responsiveness to change, we mean a responsiveness on all three organizational levels of abstraction, i.e.\ the individual, the team, and the system \citep{gren2017learning}. These levels are of course intertwined and have a complex interplay.


\section{How every other aspect of agility relates to the core}\label{sec:life} 
\subsection{The agile values and principles}
We will start with the agile manifesto and work our way through the agile principles and show how everything in the agile concept can be related to responsiveness to change. We will use the updated version of the manifesto created by Henrik Kniberg on his blog (http://blog.crisp .se/author/henrikkniberg) where he replaced some software-specific words with more general equivalents:

\begin{itemize}
\item Individuals and interactions over processes and tools.
\item Working solutions over comprehensive documentation.
\item Customer collaboration over contract negotiation.
\item Responding to feedback over following a plan.
\end{itemize}

Processes and tools need to be developed, and rigid, if they are to be accepted by employees \citep{beer1990change}. According to our experience, the reason for focusing on individuals and interactions over processes and tools is, therefore, to increase their responsiveness to change.

Extensive and overly detailed documentation takes a long time to produce, thereby reducing the flexibility of the software by increasing the cost. Additionally, documentation is often more of a formal requirement that adds little customer value. The reason for focusing on working solutions over comprehensive documentation is thus to elevate the organization's ability to adapt.

Contract negotiations are often governed by a political agenda which makes contracts between parties static and inflexible. Organizations must thus focus on customer collaboration over contract negotiation to facilitate responsiveness to change. 

Following fixed plans in a rapidly changing world are not in line with satisfying customers' changing needs. Companies need to recognize that the development of software is a knowledge-building activity in which customers and developers know more tomorrow than they do today. Creating plans and requirement specifications upfront and expecting them to remain constant throughout the development project is a utopia. Therefore, the reason for focusing on responding to feedback over following a plan is to increase the responsiveness to change. We need to note here that the fourth point in the agile manifesto was originally ``responding to change over following a plan,'' which is close to our general definition of agility. We, however, do not want this wording to be only in relation to planning, but learning about change through getting feedback.

In an attempt of making the manifesto more concrete, the authors connected a set of twelve principles to their manifesto that have been reviewed by \citet{williams}. To align the agile principles with the changes to the manifesto above, we have replaced the word \emph{software} with \emph{solutions}.
\begin{enumerate}
\item Our highest priority is to satisfy the customer through early and continuous delivery of valuable solutions. --- In order to create valuable solutions, we need to know what the customer values, and since that can change fast we need to deliver early and continuously.
\item Welcome changing requirements at the start of each iteration, even late in development; agile processes harness change for the customer's competitive advantage. --- By welcoming changes in requirements we make sure that we know when the customer changed her mind.
\item Deliver working solutions frequently, from a couple of weeks to a couple of months, with a preference for the shorter time-scale. --- In order to check if the world has changed (i.e.\ the customer or end-user), we need to deliver frequently otherwise we will not know the most recent changes.
\item The whole team, from business people through testers, must communicate and collaboratively work together throughout the project. --- Values is not only the best technical solution, therefore we need to figure out what adds the most value since this could as well be something non-technical. 
\item Build projects around empowered, motivated individuals with a shared vision of success; give them the environment and support they need, clear their external obstacles, and trust them to get the job done. --- With a shared vision, the employees can work in the same direction towards the same goal, which is hopefully aligned with adding the most value. Empowered individuals also dare to respond to change, and all the unnecessary and slow hierarchies can then be avoided. 
\item The most efficient, effective method for conveying information to and within a development team is through synchronous communication; important decisions are documented so [they] are not forgotten. --- Synchronous communication enables responsiveness to change since information then does not get stuck somewhere.
\item Valuable, high-quality solutions is the primary measure of progress at the end of each short time-boxed iteration. --- Valuable solutions must be useful solutions, which implies that we know what adds values to the customer today.
\item Agile processes promote sustainable development. The whole team should be able to maintain a reasonable work pace that includes dedicated time for exploration, visioning, re-factoring, and obtaining and responding to feedback. --- It is not possible to figure out what adds values if people are not given the opportunity to reflect, listen to feedback, and explore.
\item Continuous attention to technical excellence and good design enhances agility. --- Good designs lessens the technical debt, which increases the ability to respond to change.
\item Simplicity --the art of maximizing the amount of work not done-- is essential. --- We can only simplify our solutions if we know that they add value, otherwise we simplify waste. 
\item The best architectures, requirements, and designs emerge from self-organizing teams guided by a vision for product release. --- We need self-organizing teams because they are the only ones that can respond to change fast enough. We cannot afford to have slow decision hierarchies, and the people working on that specific solution will be able to make the most informed decision. The trick is to set the decision-making frames for teams so that they know when and which mistakes are fine to make. 
\item With each iteration, the team candidly reflects on the success of the project, feedback, and how to be more effective, then tunes and adjusts its plans and behavior accordingly. --- Reflecting on success will increase the likelihood of success in the proceeding projects. The lean aspect of efficiency also comes in here, because lean and agile are complementary concept and when we have figured out what adds value we want to deliver that efficiently. However, we want to check that continuously in order to respond to change. Since agile and lean go hand-in-hand large-scale agile frameworks (like SAFe) use the term lean\slash agile mindset and not just an agile mindset.
\end{enumerate}

\subsection{Other known definitions from the trenches}
In addition to these aspects of agility, we have essentially heard four more reoccurring definitions of what agile is by practitioners. The first one is that agile is to have less handovers in the organization. Removing silos (if they exist, and they most often do) is of course a key part in being able to respond to change, but having less handovers and follow the value flow farther in the development of, e.g.\ a component, is only to be able to respond to change, it is not the end goal. 

The second aspect is that small batch sizes is the definition of agile. Having small batch sizes is also, we argue, a means to enable responsiveness to change. With large deliveries, the world has changed too much between reality checks of what we think adds value to the customer. 

The third definition is that agility is to time box deliveries. The time boxing is mostly to assure short feedback loops to assess customer value, which also a way to be responsive to change, i.e.\ the goal is not to time box deliveries per se. Progress is measured as implemented product features and not process progress because that is how we get feedback on customer value. We do not learn about changing customer needs through internal processes improvement since optimizing internal processes is lean and not agile, i.e.\ aspects of efficiency and not effectiveness.

The catch phrase ``fail fast fix fast,'' which we have heard many times in companies we have worked with, incorporates getting feedback fast and then responding. We would only like to highlight that there is no inherent value in failing, and we could be accurate in our first try, but if that happens, agility is not an ingredient. However, we need a flexible (i.e.\ agile) development system to capture if our rapidly built initial solution was valuable or not and be able to respond to changes, and this is implemented by using feedback loops that are as short as possible.

The fourth definition is that agile is to deliver value, at least to a significant part. If we equal value delivery to being the definition of agile, we mix up agile and lean again, which, from what we have seen, is confusing to companies. Not changing a product on the way and only delivering value according to a business model implies that the requirements are stable over time and we can focus on removing waste in the process. We do not believe this has anything to do with agility, and practitioners are very much helped by making this distinction between the two.

By defining agility as responsiveness to change, we define the end goal and see all other aspects as means to that end. By doing this, it is also easy to understand that teams could be agile without anyone on the team having ever heard that term before. There are many ways of building a structure that can respond to change, but we must not confuse what we believe leads to agility and what it is. Structures and processes that lead to responsiveness to change have always existed and we must recognize those as in line with an agile approach and not something that needs to be replaced my other famous agile practices. The business of packaging agility and selling it to companies has, however, provided a case for why such approaches should be large-scale. We need to see agile as a spectrum and that companies can be more or less responsive to change. This implies that large companies can not, and should not, try to be ``fully agile.'' In any product development context with legal or any form of standards that need to be followed, such compliance will impede the agility. This is just how it is, and companies should then instead focus on where agility makes sense in their context. Embracing change as a core of agile development is far from new with this current paper, and cannot be since it has been the core of successful transformations for a long time (see \citet{beck2000epe} for example). Successful companies in a rapidly changing world will always have to balance agility and discipline \citep{boehm20}. Changing requirements has always been a challenge in software development \citep{buschmann2009learning}. In agile approaches, by contrast, the mindset is to appreciate change but this does not imply a removal of all the discipline. In other words, software engineers have come to accept the intrinsic flexibility of their own craft and finally adopted the working methods accordingly. What is interesting is that the increasingly fast changing needs in hardware development has also made hardware to behave more and more like software, but with sometimes longer feedback loops. What is very interesting is that software is now also a core part of much of the hardware development and, for example crash tests, can have very short feedback loops with the use of computer simulations. Making a waterproof distinction between hardware and software development makes less and less sense according to what we have seen lately in companies that do both.

\section{A vision for an agile transformation}\label{sec:life} 
It is easy to make the mistake of picking from all the concepts bundled into ``agility'' when creating a vision for an agile transformation. Choosing whatever solutions to real problems that we hope that agile will solve is natural. We, however, are not doing ourselves any favors through this. Adding a range of concepts to the agile transformation will only confuse the employees. One example of this is to add both responsiveness and speed to define agility when initiating organizational agile transformations. The risk is then that some employees will think that agility is speed, which we have seen many times. Speed is necessary in the agile space, but only because we quickly want to update our assumption about the world and if it has changed recently. A company might want faster internal development and remove waste, but that should not be a part of the agile transformation then. 

\begin{figure*}
\centerline{\includegraphics[width=110mm]{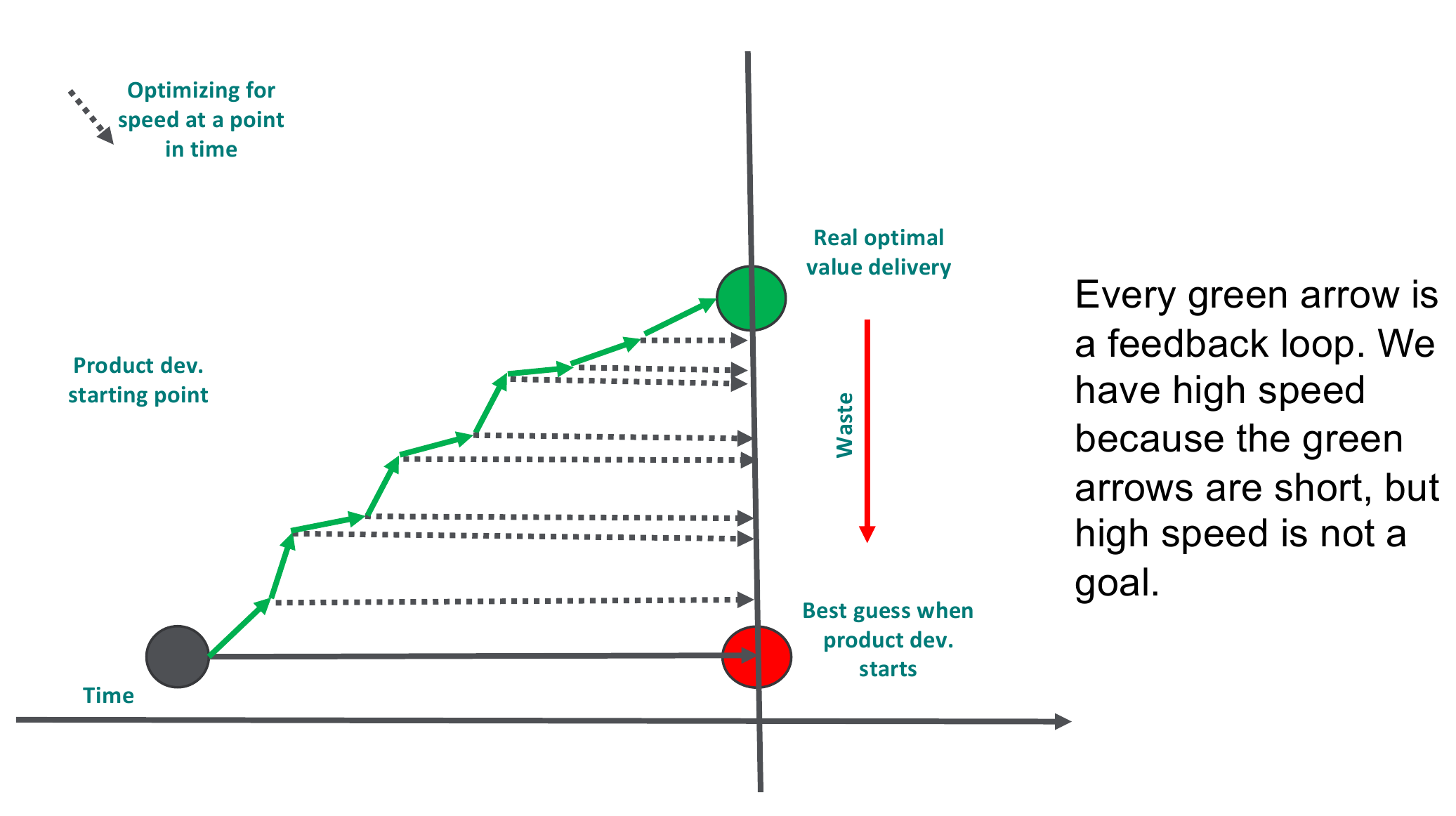}}
\caption{Optimizing for speed instead of responsiveness to change.}
\label{fig:speed}
\end{figure*}

If we confuse agility for high internal production speed and start optimizing for that, we will soon make decisions based on that speed has value in itself (see Figure~\ref{fig:speed}). We need to deploy ``sub-optimal'' solutions to collect feedback fast for us to be agile and learn what adds value, but value must have priority over speed, if we want to be agile. As we mentioned previously, agility is in relation to effectiveness (doing the right things) and lean is in relation to efficiency (doing things right). These core differences need to be made explicit when implementing the approaches. Just because lean and agile share practices, for example continuous improvement, does not imply that the core ideas behind this practice fully overlap. Instead, we view the shared practices as layers on top of, but separate from, the core intentions.

\subsection{Other confusions caused by the lack of a definition}
It is confusing that, for example, SAFe is build on four different paradigms, which includes Agile principles and methods, Lean and systems thinking, product development flow practices, and Lean processes (https://www.scaledagileframework.com/safe-lean-agile-principles/), but many practitioners think the SAFe framework is agility and react negatively when for example value stream mapping is introduced as an agile practice when they used it for many years in lean. We have also met companies that choose between agile and SAFe principles, which has devastating effects since it causes more confusion between agile and lean. It is also illogical since SAFe is partially built on the agile principles. One of the most severe negative effects of this choice is that the SAFe principles do not have any guidance for the mechanisms that need to be in place in the agile teams.

Another common resistance lately has been that agile is a software process and does not work for hardware development. If we define agility as responsiveness to change we can circumvent that resistance and also acknowledge what parts of existing hardware development processes that map very well into the agile paradigm. Many large companies have software and hardware developers even in the same team so the collaboration and the definition of a common improvement purpose are essential. If the opposite is done, i.e.\ to implement the SAFe framework top-down on hardware teams, this tends to trigger a lot of sound resistance. Again, if focus is on responsiveness to change, more people understand that focusing on structure as the end goal, that is, the ceremonies, is not what should be in focus in the agile transformation. Teams should understand the purpose of the agile transformation and be offered a smorgasbord of best practices to assess in relation to that purpose. We can then also avoid throwing the baby out with the bath water and lose context-specific knowledge essential to the companies' survival and are in line with an ability to respond to change. Research have recently confirmed many of our found challenges \citep{conboy2019implementing}, but none of the reports have suggested more focus on responsiveness to change as one of the remedies.

We have also met many teams and managers who do not understand how building teams maps into the agile idea of implementing features from a backlog. This is, of course, devastating since self-organizing teams with a mandate to make decisions within their own expertise is an absolute key to obtaining responsiveness to change (as also stated in the agile principle 11).

As mentioned, a common misconception of agility in practice is that it only means a team-based workplace. However, teams are a core part of implementing responsiveness to change, but it is not the definition of agility either. We have previously shown that the dynamics of an ideal agile team largely overlap with that of a mature team from a psychological perspective \citep{grenjss2,gren2019agile}. In our experience, the structural part is the easy part for an agile transformation, but if a company focuses more on building mature teams from a psychological perspective, the teams' responsiveness to change increases dramatically.

One of the most severe impediments we have seen in getting the teams to mature, is hierarchical ``command and control'' leadership. But the idea that an agile leader should only have a process facilitating leadership style does not make sense either. What makes sense is to train managers in that they are there for the teams and not the other way around, because leadership in old and large organization are often without such a mindset. The problem is that a more consultative leadership style is only appropriate for mature teams where they have integrated well over time, which means they must have navigated through forming, storming, and norming phases \citep{tuckman}. A completely newly formed team where no members have ever met before, a more laissez-faire leadership approach could be devastating if no natural leader steps in at an early point in time. Such teams are concerned with dependency and inclusion, and therefore need directive and clear leadership that provides the team with structure and direction to move forward \citep{wheelan2003}. It is critical, though, that the directive leadership stops when the team is mature enough to self-organize and share the leadership function (i.e.\ the drive and initiative) among its members. The agile transformation, therefore, needs to focus on guiding teams towards high levels of team maturity so that they can act independently within the set frames, and leaders must serve the teams with what they need in order to become effective. Without helping teams to mature over time, they will have difficulty building an ability to respond to changes in their development ecosystem.

We have been involved in training more than 200 teams in team development using the model by \citet{wheelan2003}, some of which have been evaluated and published recently \citet{gren2019perceived}, and also trained more than 250 team managers in supporting teams differently depending on their maturity. Some of the effects we have seen are that teams reorganize themselves across departments because the realize that they can deliver more value in that way, managers back off and help each other to back off when teams are ready to lead themselves, the teams get aware of subgroups within the teams without a common goal, and they understand how new members need to be on-boarded into their teams from a collaborative perspective. One big issue that remains is that some managers fall back into old behavior when the pressure increases on the teams' deliveries. The cultural change is, of course, what takes time and demands perseverance. 

Teams must be given the opportunity to become mature before they can be expected to self-organize and efficiently respond to change, which is not done in a day. In connection to this, realistic expectations on teams' performance must also be accounted for. According to \citet{wheelan2003}, reaching the most mature stage takes time. Their study on 114 teams showed that it takes, on average, 6.5 months to transition from the initial state (stage I) to the final stage (stage IV). This implies that organizations should not expect agile transformations to produce self-organized, empowered, agile, and high performing teams for several months, but it is worth the wait and effort. Building trust among team members simply takes time. If we want highly intelligent agile teams that are responsive to change, we must also enable them to mature, which of course also implies that they need to be as stable as possible over time.

\subsection{The effects of defining agility as responsiveness to change}

By defining agility as responsiveness to change, we think, and have seen, that agile transformations get easier. It provides clarity by reducing confusion in the organization. The organizations that make the concept of agility concrete (i.e.\ by defining what it means to them as their case of how to get higher responsiveness to change) have a more natural way of aligning the company and give a distinct direction to the transformation. This is particularly important in companies that use a scaled agile approach in which multiple teams need to collaborate. If the interpretation of the agile transformation needed is left to each team, the risk is high that the teams' interpretations differ, making it hard for them to collaborate efficiently.

By defining the core of the agile transformation, employees and managers alike can measure any changes against their potential increase in responsiveness, asking themselves ``Does this increase our responsiveness to change?'' If the answer is yes, they have a case and can defend that decision. We have met many teams that think the agile transformation only brought more hierarchies of decision-making and more overhead and waste to their process. By contextualizing the agile transformation as changes towards more responsiveness to change for the entire development system, it becomes much clearer why some teams might have to adjust their own process to obtain more overall agility even if it might hamper their own deliveries somewhat. Focusing on responsiveness to change also prevents them from thinking that agile is a set of concrete practices or simply to organize work in teams. There could be other reasons for using agile practices and arrange work in teams, such as obtaining transparency and increase job satisfaction. There could also be reason for other changes, like mapping value flows, however, the reasons behind organizational changes should be honest and explicit, otherwise people will notice and remain skeptical and confused.

Defining agility as responsiveness to change also implies that the most extreme version of agility is not reasonable for most companies, but it makes a lot of sense to discuss what contributes to agility and what hinders agility at the company. Some of the forces against agility are necessary for many companies; such as quality standards or legal and financial requirements, but one should not think they contribute to high responsiveness to change, if they do not. After listing all the forces fore and against agility, using a force field analysis approach \citep{lewin1939field} for example, one can then work on the aspects that are actually possible to change towards more agility today. 

We have recently tried deploying the agile definition as responsiveness to change to around 50 agile teams (including managers) in Sweden and 7 in China. Both team members and managers have found it very helpful to work with this definitions of agility since it makes the overall purpose of the agile transformation clear. Also, since the framework used in this particular case is built on both lean and agile ideas, they found it very useful to separate the two. Companies with a history of implementing lean, for e.g.\ manufacturing, need to know what is new in the agile space and why it is irrelevant to distinguish between hardware and software when they are two essential parts of the same system. Companies have always focused on delivering value, high speed, self-organizing teams, etc.\ so it really helps in convincing them about the missing piece that agile adds.

\section{Conclusion}
This paper set out to define agility, envisioning that it will bring clarity and direction to agile transformations. Through research and practical experience, we have found that agile transformations would be easier if agility was defined as responsiveness to change. This finding is an essential contribution to both researchers and practitioners in the agile space since they need to agree on the definition of the concept to efficiently study or apply it.

\bibliographystyle{ACM-Reference-Format}
\bibliography{references}  


\begin{thebibliography}{00}


\ifx \showCODEN    \undefined \def \showCODEN     #1{\unskip}     \fi
\ifx \showDOI      \undefined \def \showDOI       #1{{\tt DOI:}\penalty0{#1}\ }
  \fi
\ifx \showISBNx    \undefined \def \showISBNx     #1{\unskip}     \fi
\ifx \showISBNxiii \undefined \def \showISBNxiii  #1{\unskip}     \fi
\ifx \showISSN     \undefined \def \showISSN      #1{\unskip}     \fi
\ifx \showLCCN     \undefined \def \showLCCN      #1{\unskip}     \fi
\ifx \shownote     \undefined \def \shownote      #1{#1}          \fi
\ifx \showarticletitle \undefined \def \showarticletitle #1{#1}   \fi
\ifx \showURL      \undefined \def \showURL       #1{#1}          \fi
\providecommand\bibfield[2]{#2}
\providecommand\bibinfo[2]{#2}
\providecommand\natexlab[1]{#1}
\providecommand\showeprint[2][]{arXiv:#2}

\bibitem[\protect\citeauthoryear{Beck}{Beck}{2000}]%
        {beck2000epe}
\bibfield{author}{\bibinfo{person}{Kent Beck}.}
  \bibinfo{year}{2000}\natexlab{}.
\newblock \bibinfo{booktitle}{{\em Extreme Programming explained : embrace
  change}}.
\newblock \bibinfo{publisher}{Addison-Wesley}, \bibinfo{address}{Reading,
  Mass.}
\newblock


\bibitem[\protect\citeauthoryear{Beer, Eisenstat, and Spector}{Beer
  et~al\mbox{.}}{1990}]%
        {beer1990change}
\bibfield{author}{\bibinfo{person}{Michael Beer}, \bibinfo{person}{Russell~A
  Eisenstat}, {and} \bibinfo{person}{Bert Spector}.}
  \bibinfo{year}{1990}\natexlab{}.
\newblock \showarticletitle{Why change programs don't produce change}.
\newblock \bibinfo{journal}{{\em Harvard Business Review\/}}
  \bibinfo{volume}{1}, \bibinfo{number}{6} (\bibinfo{year}{1990}).
\newblock


\bibitem[\protect\citeauthoryear{Boehm and Turner}{Boehm and Turner}{2003}]%
        {boehm20}
\bibfield{author}{\bibinfo{person}{Barry Boehm} {and} \bibinfo{person}{Richard
  Turner}.} \bibinfo{year}{2003}\natexlab{}.
\newblock \bibinfo{booktitle}{{\em Balancing agility and discipline: {A} guide
  for the perplexed}}.
\newblock \bibinfo{publisher}{Addison-Wesley}, \bibinfo{address}{Boston}.
\newblock
\showISBNx{0-321-18612-5 (alk. paper)}


\bibitem[\protect\citeauthoryear{Buschmann}{Buschmann}{2009}]%
        {buschmann2009learning}
\bibfield{author}{\bibinfo{person}{Frank Buschmann}.}
  \bibinfo{year}{2009}\natexlab{}.
\newblock \showarticletitle{Learning from failure, part 1: Scoping and
  requirements woes}.
\newblock \bibinfo{journal}{{\em IEEE software\/}} \bibinfo{volume}{26},
  \bibinfo{number}{6} (\bibinfo{year}{2009}), \bibinfo{pages}{68--69}.
\newblock


\bibitem[\protect\citeauthoryear{Chow and Cao}{Chow and Cao}{2008}]%
        {chow2008survey}
\bibfield{author}{\bibinfo{person}{Tsun Chow} {and} \bibinfo{person}{Dac-Buu
  Cao}.} \bibinfo{year}{2008}\natexlab{}.
\newblock \showarticletitle{A survey study of critical success factors in agile
  software projects}.
\newblock \bibinfo{journal}{{\em Journal of systems and software\/}}
  \bibinfo{volume}{81}, \bibinfo{number}{6} (\bibinfo{year}{2008}),
  \bibinfo{pages}{961--971}.
\newblock


\bibitem[\protect\citeauthoryear{Collier}{Collier}{2011}]%
        {Collier2011aaa}
\bibfield{author}{\bibinfo{person}{Ken~W. Collier}.}
  \bibinfo{year}{2011}\natexlab{}.
\newblock \bibinfo{booktitle}{{\em Agile analytics: A value-driven approach to
  business intelligence and data warehousing}}.
\newblock \bibinfo{publisher}{Addison-Wesley}, \bibinfo{address}{Upper Saddle
  River, NJ}.
\newblock


\bibitem[\protect\citeauthoryear{Conboy and Carroll}{Conboy and
  Carroll}{2019}]%
        {conboy2019implementing}
\bibfield{author}{\bibinfo{person}{Kieran Conboy} {and} \bibinfo{person}{Noel
  Carroll}.} \bibinfo{year}{2019}\natexlab{}.
\newblock \showarticletitle{Implementing large-scale agile frameworks:
  {C}hallenges and recommendations}.
\newblock \bibinfo{journal}{{\em IEEE Software\/}} \bibinfo{volume}{36},
  \bibinfo{number}{2} (\bibinfo{year}{2019}), \bibinfo{pages}{44--50}.
\newblock


\bibitem[\protect\citeauthoryear{Gren}{Gren}{2017}]%
        {gren2017learning}
\bibfield{author}{\bibinfo{person}{Lucas Gren}.}
  \bibinfo{year}{2017}\natexlab{}.
\newblock \showarticletitle{Learning more from crossing levels: Investigating
  agility at three levels of the organization}. In \bibinfo{booktitle}{{\em
  2017 International Conference on Computational Science and Computational
  Intelligence (CSCI)}}. IEEE, \bibinfo{pages}{1035--1038}.
\newblock


\bibitem[\protect\citeauthoryear{Gren, Goldman, and Jacobsson}{Gren
  et~al\mbox{.}}{ressa}]%
        {gren2019agile}
\bibfield{author}{\bibinfo{person}{Lucas Gren}, \bibinfo{person}{Alfredo
  Goldman}, {and} \bibinfo{person}{Christian Jacobsson}.} \bibinfo{year}{In
  Press}\natexlab{a}.
\newblock \showarticletitle{Agile ways of working: {A} team maturity
  perspective}.
\newblock \bibinfo{journal}{{\em Journal of Software: Evolution and Process\/}}
  (\bibinfo{year}{In Press}).
\newblock


\bibitem[\protect\citeauthoryear{Gren, Goldman, and Jacobsson}{Gren
  et~al\mbox{.}}{ressb}]%
        {gren2019perceived}
\bibfield{author}{\bibinfo{person}{Lucas Gren}, \bibinfo{person}{Alfredo
  Goldman}, {and} \bibinfo{person}{Christian Jacobsson}.} \bibinfo{year}{In
  Press}\natexlab{b}.
\newblock \showarticletitle{The perceived effects of group developmental
  psychology training on agile software development teams}.
\newblock \bibinfo{journal}{{\em IEEE Software\/}} (\bibinfo{year}{In Press}).
\newblock


\bibitem[\protect\citeauthoryear{Gren, Torkar, and Feldt}{Gren
  et~al\mbox{.}}{2017}]%
        {grenjss2}
\bibfield{author}{\bibinfo{person}{L Gren}, \bibinfo{person}{R Torkar}, {and}
  \bibinfo{person}{R Feldt}.} \bibinfo{year}{2017}\natexlab{}.
\newblock \showarticletitle{Group development and group maturity when building
  agile teams: {A} qualitative and quantitative investigation at eight large
  companies}.
\newblock \bibinfo{journal}{{\em The Journal of Systems and Software\/}}
  \bibinfo{volume}{124} (\bibinfo{year}{2017}), \bibinfo{pages}{104--119}.
\newblock
\showDOI{%
\url{http://dx.doi.org/10.1016/j.jss.2016.11.024}}


\bibitem[\protect\citeauthoryear{Laanti, Simil{\"a}, and Abrahamsson}{Laanti
  et~al\mbox{.}}{2013}]%
        {laanti2013definitions}
\bibfield{author}{\bibinfo{person}{Maarit Laanti}, \bibinfo{person}{Jouni
  Simil{\"a}}, {and} \bibinfo{person}{Pekka Abrahamsson}.}
  \bibinfo{year}{2013}\natexlab{}.
\newblock \showarticletitle{Definitions of agile software development and
  agility}.
\newblock In \bibinfo{booktitle}{{\em Systems, Software and Services Process
  Improvement}}. \bibinfo{publisher}{Springer}, \bibinfo{pages}{247--258}.
\newblock


\bibitem[\protect\citeauthoryear{Lenberg, Feldt, and Wallgren~Tengberg}{Lenberg
  et~al\mbox{.}}{2019}]%
        {lenberg2}
\bibfield{author}{\bibinfo{person}{Per Lenberg}, \bibinfo{person}{Robert
  Feldt}, {and} \bibinfo{person}{Lars~G{\"o}ran Wallgren~Tengberg}.}
  \bibinfo{year}{2019}\natexlab{}.
\newblock \showarticletitle{Misaligned values in software engineering
  organizations}.
\newblock \bibinfo{journal}{{\em Journal of Software: Evolution and Process\/}}
  \bibinfo{volume}{31}, \bibinfo{number}{3} (\bibinfo{year}{2019}),
  \bibinfo{pages}{e2148}.
\newblock


\bibitem[\protect\citeauthoryear{Lewin}{Lewin}{1939}]%
        {lewin1939field}
\bibfield{author}{\bibinfo{person}{Kurt Lewin}.}
  \bibinfo{year}{1939}\natexlab{}.
\newblock \showarticletitle{Field theory and experiment in social psychology:
  {C}oncepts and methods}.
\newblock \bibinfo{journal}{{\em American journal of sociology\/}}
  \bibinfo{volume}{44}, \bibinfo{number}{6} (\bibinfo{year}{1939}),
  \bibinfo{pages}{868--896}.
\newblock


\bibitem[\protect\citeauthoryear{Petersen}{Petersen}{2011}]%
        {isleanagile}
\bibfield{author}{\bibinfo{person}{Kai Petersen}.}
  \bibinfo{year}{2011}\natexlab{}.
\newblock \showarticletitle{Is lean agile and agile lean? {A} comparison
  between two software development paradigms}.
\newblock In \bibinfo{booktitle}{{\em Modern software engineering concepts and
  practices: {A}dvanced approaches}}, \bibfield{editor}{\bibinfo{person}{Ali~H.
  Dogru} {and} \bibinfo{person}{Veli Bicer}} (Eds.). \bibinfo{publisher}{IGI
  Global}, \bibinfo{address}{Hershey, PA, US}, \bibinfo{pages}{19--46}.
\newblock


\bibitem[\protect\citeauthoryear{Tuckman and Jensen}{Tuckman and
  Jensen}{1977}]%
        {tuckman}
\bibfield{author}{\bibinfo{person}{Bruce Tuckman} {and} \bibinfo{person}{Mary
  Jensen}.} \bibinfo{year}{1977}\natexlab{}.
\newblock \showarticletitle{Stages of small-group development revisited}.
\newblock \bibinfo{journal}{{\em Group \& Organization Management\/}}
  \bibinfo{volume}{2}, \bibinfo{number}{4} (\bibinfo{year}{1977}),
  \bibinfo{pages}{419--427}.
\newblock


\bibitem[\protect\citeauthoryear{Wheelan, Davidson, and Tilin}{Wheelan
  et~al\mbox{.}}{2003}]%
        {wheelan2003}
\bibfield{author}{\bibinfo{person}{Susan Wheelan}, \bibinfo{person}{Barbara
  Davidson}, {and} \bibinfo{person}{Felice Tilin}.}
  \bibinfo{year}{2003}\natexlab{}.
\newblock \showarticletitle{Group Development Across Time: {R}eality or
  Illusion?}
\newblock \bibinfo{journal}{{\em Small group research\/}} \bibinfo{volume}{34},
  \bibinfo{number}{2} (\bibinfo{year}{2003}), \bibinfo{pages}{223--245}.
\newblock


\bibitem[\protect\citeauthoryear{Williams}{Williams}{2012}]%
        {williams}
\bibfield{author}{\bibinfo{person}{Laurie Williams}.}
  \bibinfo{year}{2012}\natexlab{}.
\newblock \showarticletitle{What agile teams think of agile principles}.
\newblock \bibinfo{journal}{{\it Commun. ACM}} \bibinfo{volume}{55},
  \bibinfo{number}{4} (\bibinfo{year}{2012}), \bibinfo{pages}{71--76}.
\newblock


\bibitem[\protect\citeauthoryear{Womack and Jones}{Womack and Jones}{2003}]%
        {Womack2003ltb}
\bibfield{author}{\bibinfo{person}{James~P. Womack} {and}
  \bibinfo{person}{Daniel~T. Jones}.} \bibinfo{year}{2003}\natexlab{}.
\newblock \bibinfo{booktitle}{{\em Lean thinking: Banish waste and create
  wealth in your corporation}}.
\newblock \bibinfo{publisher}{Free Press Business}, \bibinfo{address}{London}.
\newblock


\end{thebibliography}

\end{document}